\documentclass[superscriptaddress, nofootinbib, prd, twocolumn, letterpaper]{revtex4}[12pt]
\usepackage{graphicx}
\graphicspath{{figs/}}
\usepackage {amsmath, amssymb, rotating}
\usepackage{hyperref}
\usepackage[usenames,dvipsnames]{color}
\usepackage{mdwlist} 
\usepackage{slashed}
\usepackage{verbatim}
\usepackage{mathrsfs}

\usepackage{epsfig,graphics}
\usepackage{epstopdf}

\DeclareMathOperator{\sinc}{sinc}

\begin{document}

\title{A Resonant Mode for Gravitational Wave Detectors based on Atom Interferometry}

\author{Peter W. Graham}
\affiliation{Stanford Institute for Theoretical Physics, Department of Physics, Stanford University, Stanford, CA 94305}

\author{Jason M. Hogan}
\affiliation{Department of Physics, Stanford University, Stanford, CA 94305}

\author{Mark A. Kasevich}
\affiliation{Department of Physics, Stanford University, Stanford, CA 94305}

\author{Surjeet Rajendran}
\affiliation{Berkeley Center for Theoretical Physics, Department of Physics, University of California, Berkeley, CA 94720}

\begin{abstract}
We describe an atom interferometric gravitational wave detector design that can operate in a resonant mode for increased sensitivity.  By oscillating the positions of the atomic wavepackets, this resonant detection mode allows for coherently enhanced, narrow-band sensitivity at target frequencies.  The proposed detector is flexible and can be rapidly switched between broadband and narrow-band detection modes.  For instance, a binary discovered in broadband mode can subsequently be studied further as the inspiral evolves by using a tailored narrow-band detector response.  In addition to functioning like a lock-in amplifier for astrophysical events, the enhanced sensitivity of the resonant approach also opens up the possibility of searching for important cosmological signals, including the stochastic gravitational wave background produced by inflation.  We give an example of detector parameters which would allow detection of inflationary gravitational waves down to $\Omega_\text{GW} \sim 10^{-14}$ for a two satellite space-based detector.
\end{abstract}

\date{\today}
\maketitle

\section{Introduction}

LIGO has opened a new window into the universe by successfully launching the era of gravitational wave astronomy \cite{TheLIGOScientific:2016qqj}. The LIGO detector is sensitive to gravitational waves with frequencies $\gtrapprox~10$~Hz~\cite{TheLIGOScientific:2016agk}.  It is also of interest to probe the gravitational wave spectrum in the frequency band 0.1~Hz - 10~Hz \cite{SatyaSchutz}. Inspiralling black hole and neutron star binaries have to travel through this lower frequency band before entering LIGO's band \cite{Sesana2016}. Continuous observation of these inspirals may allow better estimation of binary parameters, potentially enabling their use as  standard sirens for cosmological measurements \cite{sirens}. Further, extremal black holes observed well before their final coalescence can  constrain  the existence of ultra-light bosonic particles such as axions \cite{Arvanitaki:2014wva}. In addition to sources that are observable at LIGO, the mergers of intermediate mass black holes \cite{MillerColbert} (that do not make it to LIGO's band) occur in the frequency band 0.1 Hz - 10 Hz. This band is also expected to be relatively free of unresolvable astrophysical gravitational wave backgrounds, making it an optimal choice to search for cosmological sources of stochastic gravitational waves such as those expected from cosmic inflation \cite{Marx:2011hf, CoughlinHarms, TheLIGOScientific:2009Nature}. 

Atom interferometers have the potential to probe this spectrum in terrestrial and satellite-based experiments \cite{Hogan:2015xla, Graham:2012sy, Hogan:2010fz, Dimopoulos:2008sv, Dimopoulos:2007cj, Chaibi2016}.  These proposals are motivated by recent, demonstrated advances in atom technology such as the ability to perform large momentum transfer beamsplitters \cite{Robins80hk} to coherently separate atomic wave-packets by $\sim$ 54 cm \cite{54cm}, delta kick cooling \cite{MuntingaBEC} of atom ensembles to pK temperatures \cite{picokelvin} and the development of optical clock metrology \cite{OpticalClock}.  Additionally, the original detector concept required two-photon transitions which introduce laser noise into the measurement (see e.g.~\cite{Dimopoulos:2008sv}).  Using single photon transitions was known to preserve the needed `accelerometer' signal and remove laser noise (see e.g.~\cite{Dimopoulos:2008hx}) but is not sensitive enough to detect gravitational waves (see e.g.~\cite{Yu:2010ss}).  This problem is solved by a new technique which allows large momentum transfer beamsplitters and hence a sensitive gravitational wave detector without laser noise \cite{Graham:2012sy}.  Thus atom interferometry shows significant potential for gravitational wave detection.

In the detector scheme considered in this paper \cite{Hogan:2015xla, Graham:2012sy}, two widely separated atom interferometers  (essentially optical clocks) are run using common lasers, where the lasers drive single-photon transitions in the atoms. Each interferometer can be thought of as precisely comparing the time kept by the laser's clock (the laser's phase), and the time kept by the atom's clock (the atom's phase). A passing gravitational wave changes the normal flat space relation between these two clocks by a factor proportional to the distance between them. This change oscillates in time with the frequency of the gravitational wave.  This is the signal that can be measured with an atom interferometer.  The signal in an individual interferometer is masked by laser phase, frequency  and platform noise. However, since the same laser beams are used to operate both interferometers, their differential phase contains the gravitational wave signal with severely suppressed laser noise. This differential signal enables the operation of gravitational wave detectors with a single baseline, as opposed to the two (or more) baselines demanded by conventional optical interferometers. In addition, this approach also exploits the fact that the atom ensembles are inherently mechanically isolated from the environment as they free-fall during the optical interrogation sequence \cite{Hogan:2010fz, Dimopoulos:2008sv, Dimopoulos:2007cj}. We note that instead of single-photon optical clock transitions, the interferometer can also be operated using conventional two-photon Bragg/Raman transitions wherein the atoms serve as a phase reference to the lasers used to drive the interferometers. In this case, since the lasers are the only available clock, cancellation of laser frequency noise requires the operation of two (or more) baselines \cite{Hogan:2010fz, Dimopoulos:2008sv, Dimopoulos:2007cj} or an additional optical frequency reference (e.g., an optical lattice clock \cite{OpticalClock} and frequency comb \cite{Udem2002}). The resonant scheme developed in this paper can be applied both to the optical clock and two photon Bragg/Raman atom interferometer configurations.

The conventional light-pulse atom interferometer configuration based on a $\pi/2$-$\pi$-$\pi/2$ pulse sequence is maximally sensitive to gravitational waves whose frequency $\omega$ is $\pi/T$, where $T$ is the interrogation time of the interferometer (the time interval between successive pulses).  When this condition is satisfied, the gravitational wave completes a full cycle in time $2T$, resulting in the maximum possible change of the baseline's proper length during the interferometer.  For frequencies $\omega \gg 1/T$,  the wave oscillates many times during $T$. Since the arms of the interferometer remain fixed during these rapid oscillations, the effects of the wave largely cancels, except for phase shifts  accrued over a single period $\sim 1/\omega$ of the wave. The instrument's mechanical design fixes the maximum interrogation time $T_{\text{max}}$  \cite{Dimopoulos:2008sv, Dimopoulos:2007cj} and sets the  lowest frequency accessible to the detector. In this paper, we show that for frequencies $\omega \gg 1/T_{\text{max}}$, the pulse sequences used to drive the interferometer can be changed to {\it resonantly} enhance the sensitivity of the detector.  Hence, the same instrument can be switched in real time to operate in a more sensitive, resonant mode for frequencies $\omega \gg 1/T_{\text{max}}$, albeit at the expense of bandwidth. We show that this narrow band resonant mode enhances the sensitivity of the detector to both coherent signals (such as inspiralling binaries) and stochastic sources.

\section{Resonant Mode}

\begin{figure}[t]
	\begin{center}
		\includegraphics[width=\columnwidth]{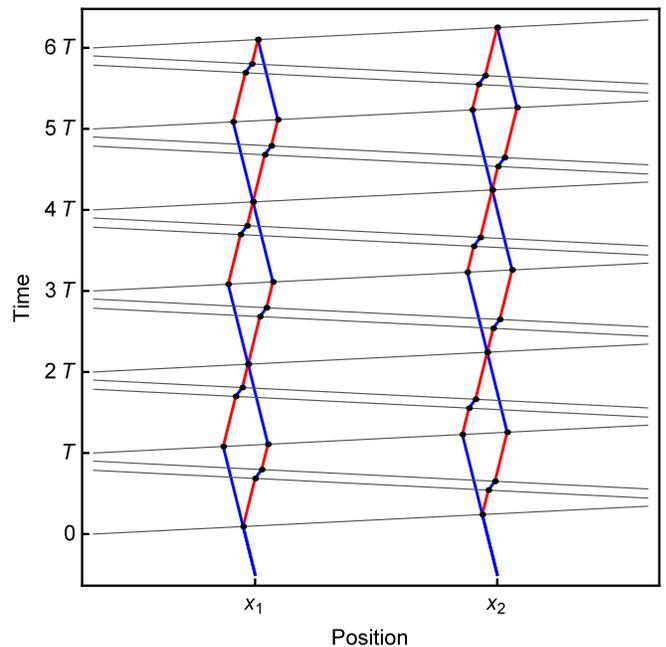}
		\caption{
			A space-time diagram of an example resonant atom interferometer detector sequence.  The detector consists of two atom interferometers, one at position $x_1$ and the other at position $x_2$, where $x_2 = x_1 + L$ and $L$ is the baseline.  The spacetime trajectories of the atoms are shown in blue for the ground state and red for the excited state. Pulses of light (thin black lines) are sent back and forth from each end of the baseline and interact with the atoms (interactions shown as black dots), transferring momentum to the atoms and changing their internal state.  Whether or not an interaction occurs is controlled by matching the frequency of the light pulses to the Doppler shift of the atoms.  The sequence shown consists of two single-photon transitions for each atom optic ($n=2$, $2\hbar k$ momentum transferred) and a resonant enhancement of $Q=3$ (three diamonds).  The interrogation time $T$ depicted here is near the limit for this baseline $T \approx 2(n-1) L/c$, so the gaps between the sequential laser pulses are small.
		}
		\label{fig:many diamonds}
	\end{center}
\end{figure}

The interferometer can be run in a resonant mode by using the pulse sequence $\pi/2 - \pi -  \dots - \pi - \pi/2$  (see Fig.~\ref{fig:many diamonds}) with $Q$ $\pi$ pulses instead of the standard, broadband $\pi/2 - \pi - \pi/2$ pulse sequence. The pulses are equally spaced in time by $T \lessapprox  T_{\text{max}}/Q$. Gravitational waves with frequencies $\omega = \pi/T$ oscillate by half a cycle between the pulses (e.g., from crest to trough). Unlike the broadband case, the phase differences caused by subsequent oscillations continually add in the resonant sequence, since the series of $\pi$ pulses periodically swap the arms of the interferometer.  Such pulse sequences have been previously explored in the context of optical clocks, where they have been used to characterize laser phase noise \cite{Ye_dynamic_decoupling}.  They have also been recently proposed in the context of gravitational wave detection using optical lattice clocks \cite{Shimon}.

\subsection{Detector response function}

\begin{figure}[t]
	\begin{center}
		\includegraphics[width=\columnwidth]{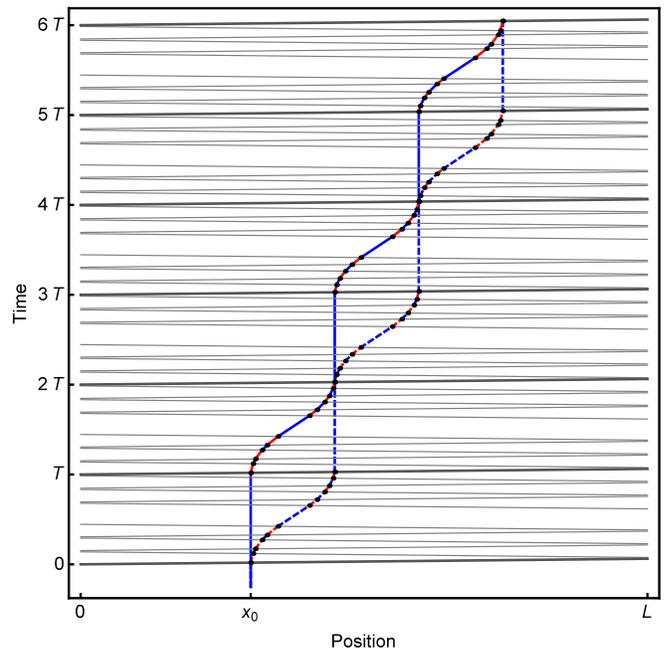}
		\caption{
			A space-time diagram of an LMT enhanced resonant atom interferometer sequence.  This pulse sequence uses an LMT enhancement of $n=6$ ($6\hbar k$ momentum transfer per LMT atom optic) and a resonant enhancement of $Q=3$ (three diamonds).  The dashed curve is the atom trajectory of the upper arm of the interferometer and the solid curve is the lower arm.  Pulses of light are shown as gray lines traveling in alternating directions from each end of the baseline.  The midpoint of each LMT $\pi$-pulse sequence is indicated with a dark gray photon path, while the $(2n-1)$ axillary $\pi$ pulses are shown in light gray.  The first and last pulse sequences are the LMT $\frac{\pi}{2}$ pulses and consist of $(n-1)$ axillary $\pi$ pulses (light gray) and one $\frac{\pi}{2}$ pulse (dark gray).  As in Fig.~\ref{fig:many diamonds}, the atom ground state is shown in blue and the excited state is red, and black dots indicate atom-laser interaction points.  For simplicity, only one LMT resonant interferometer is shown here (at position $x_0$), but in the detector configuration there would be two interferometers at different positions, as in Fig.~\ref{fig:many diamonds}.  For the parameters chosen here, the light travel time of the $n$ pulses in each LMT atom optic is comparable to the interrogation time $T$ of the interferometer, so the atom trajectories exhibit substantial curvature.  Also, the recoil velocity $v_r$ has been greatly exaggerated to show detail, but in any real detector we always have $v_r T \ll L$. 
		}
		\label{fig:lmtAI}
	\end{center}
\end{figure}

As usual, we consider a configuration consisting of two atom interferometers separated by a baseline of length $L$ that are driven by a sequence of common laser pulses.  In the following we assume the interferometers are based on single-photon transitions, although as mentioned above a resonant sequence can similarly be implemented using conventional two-photon atom optics.  To take advantage of large momentum transfer (LMT) phase enhancement, each atom optics `pulse' may actually consist of $n$ closely spaced $\pi$-pulses which increase the separation of the interferometer arms (see Fig.~\ref{fig:lmtAI}).  

We use the following metric written in the TT gauge for a plane gravitational wave traveling the $z$-direction:
\begin{align}
ds^2=c^2dt^2-&dx^2-dy^2-dz^2 + h(t)(dx^2-dy^2) \label{Eq:GWmetric}
\end{align}
where $h(t) = h\cos{(\omega(t-\tfrac{z}{c})+\phi_0)}$ and $h$ is the strain amplitude of the gravitational wave, $\omega$ is the frequency, and $\phi_0$ is the initial phase.  For simplicity we assume the interferometers lie along the $x$-axis so that the detector is maximally aligned with the GW strain.  It is straightforward to generalize this to arbitrary gravitational wave polarization and angle of incidence, as is treated in \cite{Hogan:2015xla}.

To calculate the detector response, we determine the phase shift of the resonant interferometer using the methods described in previous work \cite{Dimopoulos:2008hx,Graham:2012sy,Hogan:2015xla}. The phase shift can be expressed as the sum of three contributions: the propagation phase, laser phase, and separation phase.  The laser phase cancels to first order in a single-photon gradiometer \cite{Graham:2012sy}, and the separation phase contribution to the signal can be neglected since the wavepacket separation is small compared to $L$ \cite{Hogan:2015xla}.  Thus the leading order signal comes from a difference in propagation phase between the two interferometers, which arises because the atoms on one end of the baseline spend a different amount of time in the excited state than atoms on the other end.

Figure~\ref{fig:lmtAI} shows an LMT interferometer sequence.  The interferometer is divided into segments in the ground (blue) and excited (red) states.  The duration of each segment is determined by the intersection of the light pulses (gray) with the atomic trajectories.  Using the metric Eq.~\ref{Eq:GWmetric}, a light pulse that leaves one of the laser sources at position $x_I$ and time $t_I$ will arrive at the atom at position $x$ at coordinate time \cite{Rakhmanov2009,Hogan:2015xla}
\begin{align}
\tau(x,x_I,t_I) = t_I+\tfrac{|x-x_I|}{c} + \tfrac{h}{2\omega}\Big(& \sin{\!\big( \tfrac{\omega|x-x_I|}{c} +\omega t_I +\phi_0 \big)} \nonumber \\
&- \sin{\!\big( \omega t_I +\phi_0 \big)} \Big) \label{Eq:ArrivalTimePhoton}
\end{align}
These arrival times determine the duration of the excited state segments.  Neglecting recoil effects \cite{Hogan:2015xla}, the geodesics of the atoms can be approximated as being at fixed coordinate positions $x=\text{constant}$, so the contribution to the propagation phase from the $i^\text{th}$ segment reduces to
\begin{equation}
\delta\phi_i = \frac{c}{\hbar}\int m_i ds = \frac{m_i c^2}{\hbar}\delta \tau_i
\end{equation}
where $m_i c^2$ is the mass-energy of the atom along the segment and $\delta \tau_i$ is the duration of the segment, determined by Eq.~\ref{Eq:ArrivalTimePhoton}.  For the ground state the mass-energy is $m c^2$ while for the excited state it is $m c^2+\hbar \omega_a$, where $\hbar\omega_a$ is the atomic energy level spacing.  The total propagation phase is
\begin{equation}
\Delta\Phi_\text{prop} = \sum_{\{u_i\}} \delta\phi_i - \sum_{\{l_i\}} \delta\phi_i\label{Eq:propPhase}
\end{equation}
where the sums are over the set of all upper segments $\{u_i\}$ and lower segments $\{l_i\}$ of the interferometer.

In Fig.~\ref{fig:lmtAI}, each diamond-shaped loop has duration $2T$ and there are $Q$ diamonds in total.  The emission times of the light pulses are given by $t_{q,i}\equiv t_0 + q T + i\tfrac{L}{c}$, where $0 \le q \le 2Q$ and $0\le i < n$ are integers.  Pulses with even $i$ originate from $x_I=0$ while pulses with odd $i$ originate from $x_I=L$.  During the first half of each diamond (i.e., between time $q T$ and $(q+1)T$), the upper arm of the interferometer spends $n$ intervals of time in the excited state while the lower arm remains in the ground state, and the opposite holds during the second half of each diamond.

By symmetry, each pulse at time $t_{q,i}$ that adds momentum to the arm is later followed by a pulse originating from the same $x_I$ at time $t_{q+1,-i}$ that subtracts momentum.  These pulse pairs define a set of nested intervals of duration $\Delta t_i=(t_{q+1,-i} - t_{q,i})$, where $\Delta t_{i+1} < \Delta t_i$ and for the $i^\text{th}$ and $(i+1)^\text{th}$ intervals the atom is in alternating atomic states.  Time intervals that are bracketed by pulses from $x_I=0$ add to the time in the exited state (even $i$ pulses), while intervals due to pulses from $x_I=L$ reduce that time (odd $i$ pulses).

Let $\Phi(q,x)$ be the net phase accumulated during the $q^\text{th}$ half-diamond for an atom at position $x$.  Summing over all the even $i$ intervals and subtracting the odd $i$ intervals yields a net phase of
\begin{align}
\label{Eq:halfdiamondphase}
\Phi(q,x) = \omega_a & \sum_{\substack{i=0 \\ \text{even}}}^{n-1}\Big(\tau(x,0,t_{q+1,-i})-\tau(x,0,t_{q,i})\Big) \\ \nonumber
&-\omega_a\sum_{\substack{i=0 \\ \text{odd}}}^{n-1}\Big(\tau(x,L,t_{q+1,-i})-\tau(x,L,t_{q,i})\Big) 
\end{align}
The phase for a complete diamond is given by the difference of $\Phi(q,x)$ for two arms, which by symmetry are related by a time translation by $T$.  Summing over all $Q$ diamonds in the sequence gives the total phase for an interferometer at position $x$:
\begin{equation}
\Delta\Phi(x) = \sum_{j=0}^{Q-1}\Big(\Phi(2j+1,x)-\Phi(2j,x) \Big)
\end{equation}
The detector response is the phase difference of an interferometer at $x=0$ and $x=L$, which we call the gradiometer phase: $\Delta\Phi_\text{grad}\equiv\Delta\Phi(L)-\Delta\Phi(0)$.  We define the difference in light arrival time $\Delta\tau$ at the location of the atoms at each end of the gradiometer as
\begin{align}
\Delta\tau(t) &\equiv \tau(L,x_I=0,t) - \tau(0,x_I=0,t) \nonumber \\
&= \tfrac{L}{c}+\tfrac{h}{2\omega}\Big( \sin{\!\big( \omega (t+\tfrac{L}{c}) +\phi_0 \big)} - \sin{\!\big( \omega t +\phi_0 \big)} \Big) \nonumber \\
&= -\Big(\tau(L,x_I=L,t) - \tau(0,x_I=L,t)\Big) 
\end{align}
Note that the arrival time differences for pulses originating from $x_I=0$ and $x_I=L$ are the same except for an overall minus sign.  This means that the even and odd sums of Eq.~\ref{Eq:halfdiamondphase} can be combined, yielding
\begin{align}
\Delta\Phi_\text{grad} = \omega_a \sum_{j=0}^{Q-1}\sum_{i=0}^{n-1}\Big( &\Delta\tau(t_{2j+2,-i})-\Delta\tau(t_{2j+1,i}) \nonumber \\ 
&-\Delta\tau(t_{2j+1,-i})+\Delta\tau(t_{2j,i}) \Big) 
\end{align}

\noindent The response of the detector has the form $\Delta\Phi_\text{grad}(t_0)=\Delta\phi \cos{(\omega t_0+\phi_0)}$, where $\omega t_0+\phi_0$ is the phase of the gravitational wave at time $t_0$ at the start of the pulse sequence.  The amplitude of the detector response is 
\begin{equation}
\Delta\phi= k_\text{eff} h L \frac{\sin\!{(\omega Q T)} }{\cos\!{(\omega T/2)}} \sinc\!{\big(\hspace{-0.1em}\tfrac{\omega n L}{2 c}\hspace{-0.1em}\big)}  \sin\!{\big(\hspace{-0.1em}\tfrac{\omega T}{2}\!-\!\tfrac{\omega(n-1)L}{2c}\hspace{-0.1em}\big)}\label{Eq:PhaseShift}
\end{equation}
Here $\hbar k_\text{eff}$ is the momentum transferred to the atom during the LMT beamsplitters and mirrors, where $k_\text{eff}\equiv n \omega_A/c$ for an $n$-pulse LMT sequence using an optical transition with atomic energy level spacing $\hbar \omega_A$.  The response is peaked at the resonance frequency 
$\omega_r \equiv \pi/T$ and has a bandwidth given by $\sim \omega_r/Q$.  The peak phase shift on resonance ($\omega=\omega_r$) has amplitude
\begin{equation}
\Delta\phi_\text{res}=2Q k_\text{eff} h L \sinc\!{\big(\tfrac{\omega_r n L}{2 c}\big)} \cos\!{\big(\tfrac{\omega_r (n-1) L}{2 c}\big)}\label{Eq:PeakPhaseShift}
\end{equation}
which in the low frequency limit $\omega_r \ll \tfrac{c}{n L}$ reduces to $\Delta\phi_\text{res}\approx 2 Q k_\text{eff} h L$.  As expected, the phase response shows an $n$-fold sensitivity enhancement from LMT and a $Q$-fold enhancement from operating in resonant mode.  The interferometer can be switched from broadband to resonant mode by changing the pulse sequence used to operate the device (changing $Q$). Note that the implementation of this scheme principally requires the ability to coherently transfer momenta $k_{\text{eff}}$ to the atom wavepackets $Q$ times, much like LMT interferometry. The practicalities of this are discussed in Section~\ref{Sec:Discussion}.

\subsection{Stochastic sensitivity}

This resonant strategy may give significant sensitivity to a stochastic background of gravitational waves, as can arise for example from inflation.  For such a measurement it is desirable to cross-correlate two detectors.  This could be, for example, two single-arm detectors of the type described above.  Following \cite{Christensen:1992wi} (but see also \cite{Flanagan:1993ix, Allen:1997ad}) we estimate the sensitivity (really the $95 \%$ confidence limit) to such a stochastic background as
\begin{equation}
\label{eqn: stochastic sensitivity}
\Omega_\text{GW} (f) = \frac{\pi c^2 f^3}{\rho_c G | \gamma \left( \vec{x}_1, \vec{x}_2, f \right) |} \sqrt{\frac{2}{\tau_\text{int} \Delta f}} (1.645) h_n^2 (f)
\end{equation}
where $\tau_\text{int}$ is the total averaging time of the experiment, $\gamma$ is a geometric factor taking into account the positions of the two detectors which we take equal to its maximum value $\frac{8 \pi}{5}$ (it will probably be slightly smaller in a real configuration), and $\rho_c$ is the closure density of the Universe.  Here $h_n$ is the amplitude spectral density of the strain noise in the gravitational wave detector as plotted for example in Fig.~\ref{fig:strain sensitivity}.  For the bandwidth we take $\Delta f \sim \frac{f_r}{Q}$ where $f_r=\omega_r/2\pi$ is the resonant frequency of the detector.

The sensitivity in Eqn.~\eqref{eqn: stochastic sensitivity} improves with resonance as $\Omega_\text{GW} \propto Q^{- \frac{3}{2}}$ (for a fixed resonant frequency $f_r$).  This is because the strain sensitivity $h_n$ improves linearly in $Q$, but we lose because the bandwidth drops linearly $\Delta f \propto Q^{-1}$ so we are integrating over less power at higher $Q$.  A rough, intuitive understanding of this scaling with $Q$ is as follows.  The relevant part of the stochastic signal is the power within the bandwidth $\Delta f$, so assume that in fact the entire signal is just this part of the power spectrum.  This piece of the signal power spectrum by itself is sharply peaked in frequency with a coherence time of $\left( \Delta f \right)^{-1}$.

\section{Strain Sensitivity}

\begin{figure}[t]
	\begin{center}
		\includegraphics[width=\columnwidth]{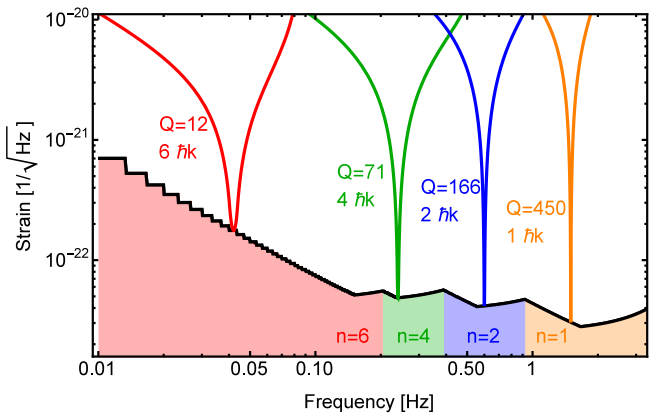}
		\caption{
			Strain sensitivity for a detector using resonantly enhanced LMT atom interferometry. The design is based on single-photon transitions in Sr and uses a heterodyne laser link over a single baseline of length $L=4.4\times 10^{7}~\text{m}$.  The atom interferometer phase noise is taken to be $\delta\phi_a = 10^{-5}~\text{rad}/\sqrt{\text{Hz}}$, and the design assumes a telescope diameter of $d=50~\text{cm}$ which is large enough to keep the contribution of photon shot noise less than $\delta\phi_a$ for all frequencies in the detection band.  The sensitivity curves for several possible pulse sequences are shown (red, green, blue, orange) with different choices for the resonant enhancement $Q$ and LMT $n$.  The black curve shows the peak on-resonance response at each frequency that can be reached with appropriate choice of $n$ and $Q$.  In all cases, the parameters of the sequence are constrained so that the total number of pulses does not exceed $n_\text{max} = 10^3$ and the total interferometer duration is less than $T_\text{max} = 300~\text{s}$.  The shaded regions below the black curve indicate the value of $n$ used in each range.  The repetition rate of the detector is $f_\text{rep} = \tfrac{\chi}{2T}$, where we assume sufficient multiplexing to reach $\chi = 10$ samples per gravitational wave period.
			}
		\label{fig:strain sensitivity}
	\end{center}
\end{figure}

Figure~\ref{fig:strain sensitivity} shows example strain sensitivities curves that can be achieved using the resonant detector mode.  We consider a specific design with a baseline of $L=4.4\times 10^{7}~\text{m}$, chosen to highlight the sensitivities that can be achieved in the $0.1~\text{Hz}$ - $1~\text{Hz}$ band.  To take advantage of this long baseline, we assume a detector design based on heterodyne laser links as described in \cite{Hogan:2015xla}.  Each of the curves (red, green, blue, orange) shows the strain sensitivity derived from Eq.~\ref{Eq:PhaseShift} for a specific choice of the tunable parameters of the detector: interferometer time $T$, resonant enhancement $Q$, and LMT enhancement $n$.  While $T$ is set by the targeted resonant frequency, the values of $Q$ and $n$ have been chosen in each case to maximize the sensitivity while respecting certain practical constraints, as described in detail below.  Any of these sensitivity curves can be realized by the same physical detector (e.g., with the same baseline, telescope, and laser design) and the response of the detector can be rapidly changed among these by changing the pulse sequence in software.  In Fig.~\ref{fig:strain sensitivity} we also assume that laser frequency noise is sufficiently common between the two interferometers that it can be neglected, as described in \cite{Hogan:2015xla,Graham:2012sy}.  The implied requirements for laser frequency noise (and other technical noise sources) are discussed in Sec.~\ref{NoiseReqs}.

The black curve in Fig.~\ref{fig:strain sensitivity} shows the peak strain sensitivity for a resonant sequence at each frequency, as given by Eq.~\ref{Eq:PeakPhaseShift} and using the same example design ($L=4.4\times 10^{7}~\text{m}$).  As such, this is not the broadband instantaneous response of the detector, but rather it shows the strain sensitivity that can be achieved at each frequency using a narrow band resonant sequence.  Once again, since the operating frequency $1/T$ can be easily changed, the detector can be scanned sequentially across this band to study signals at all frequencies.

The strain sensitivity also depends on the minimum resolvable phase shift in the interferometer.  Assuming technical sources of phase noise can be suppressed sufficiently well as a common-mode, the detector phase readout is ultimately limited by the quantum projection noise of the atoms.  Here we anticipate that the performance of the detector can benefit from ongoing advances in rapid ultracold atom preparation\cite{stellmer2013laser}, as well as from recent results\cite{hosten20dB} demonstrating metrologically significant spin squeezing.  For the curves in Fig.~\ref{fig:strain sensitivity}, we extrapolate beyond current state-of-the-art and assume a phase noise of $\delta\phi_a = 10^{-5}~\text{rad}/\sqrt{\text{Hz}}$.  This could be met with atom shot-noise-limited detection using an atom flux of $10^{10}~\text{atoms}/\text{s}$.  Alternatively, with $20~\text{dB}$ of spin squeezing\cite{hosten20dB} the same target phase noise would require an atom flux of $10^{8}~\text{atoms}/\text{s}$.

\section{Discussion} \label{Sec:Discussion}

\subsection{Detector design}

The optimal choice of parameters, for example the choice of $Q$, for an actual gravitational wave detector is complicated and depends on the signal being searched for.  A full experimental design is beyond the scope of this paper.  Here we do not attempt to describe the optimal configurations for all situations.  The parameters chosen in Figure \ref{fig:strain sensitivity} are not optimized for science reach or to avoid technical noise sources.  Here we simply describe the tool of resonant atom interferometry and several different possible uses.

One consideration is that in certain configurations there may be a tradeoff between the resonant enhancement $Q$ and the LMT enhancement $n$ since both require repeated atom-laser interactions.  For example, there is a practical limit to the total number of atom-laser interactions that are possible before an order one fraction of the atoms are not coherently transferred.  A benefit to using more LMT instead of more resonant enhancement is that while both increase the signal linearly, LMT leaves the detector broadband.  This tradeoff is discussed in more detail in Sec.~\ref{Sec:SensitivityConstraints}.

On the other hand, it is possible that atom loss is not the limiting factor.  For instance, one limit on LMT is the spatial size of the interferometer region, which can become large because of the wavepacket separation associated with LMT \cite{54cm}.  Using resonant enhancement while reducing LMT allows the interferometer region to remain smaller because the two halves of the atom separate at a smaller recoil velocity $\hbar k_\text{eff}/m$.  In order to maintain a fixed target sensitivity without resonant enhancement ($Q=1$), the product $n Q$ must be kept constant by increasing $n$. For example, in the case of the green sensitivity curve in Fig.~\ref{fig:strain sensitivity}, maintaining the same sensitivity at the target frequency with $Q=1$ would require $n=284$, resulting in a wavepacket separation of $\Delta x=\hbar k_\text{eff}T/m\sim 8~\text{m}$.  By comparison, using a resonant sequence ($n=4$ and $Q=71$) leads to a wavepacket separation of only $\Delta x\sim 10~\text{cm}$.  This tradeoff could allow a higher sensitivity interferometer to fit inside a fixed region such as a satellite.  For a satellite-based gravitational wave detector, this could be a significant advantage of using this resonant mode of operation.

\subsection{Application to cosmology and astrophysics}

\begin{figure}[t]
	\begin{center}
		\includegraphics[width=\columnwidth]{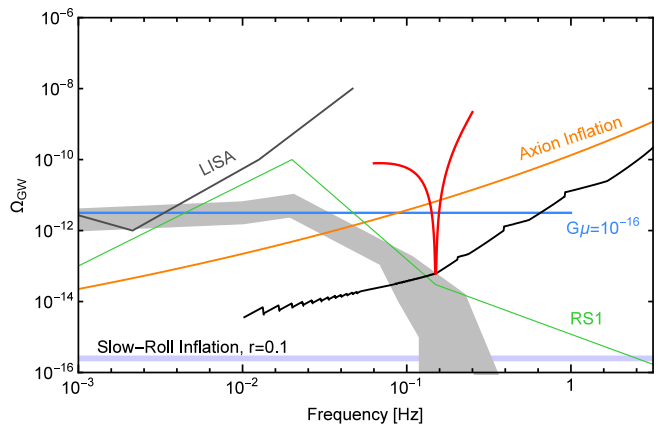}
		\caption{
			Stochastic gravitational wave sensitivity.  The black curve shows the sensitivity to $\Omega_\text{GM}$ using the same detector design as Fig.~\ref{fig:strain sensitivity}, with baseline length $L=4.4\times 10^{7}~\text{m}$,  phase readout noise $\delta\phi_a = 10^{-5}~\text{rad}/\sqrt{\text{Hz}}$, and telescope size $d=50~\text{cm}$.  This curve is constructed from Eq.~\ref{eqn: stochastic sensitivity} using the peak on-resonance response for $h_n(f)$ (the black curve from Fig.~\ref{fig:strain sensitivity}), a bandwidth of $\Delta f = f/Q$ (using the appropriate $Q$ at each $f$), and a $\tau_\text{int} = 1~\text{year}$ integration time.  As before, this is not the broadband sensitivity of the detector, but rather it shows the sensitivity that can be reached at each frequency after operating a narrow-band resonant detector there for $\tau_\text{int} = 1~\text{year}$.  To help visualize this, the red curve shows the $\Omega_\text{GM}$ sensitivity for a specific sequence ($h_n$ given by Eq.~\ref{Eq:PhaseShift}) with resonance frequency $f_r=0.15~\text{Hz}$, $Q=44$, and $6\hbar k$ LMT atom optics.  This frequency is in a range that is interesting for cosmological signals and may avoid white dwarf binary confusion noise, shown in the gray band (whose width shows an estimated uncertainty) \cite{Farmer:2003pa}.  The purple line shows an estimate for the power spectrum that could arise from GWs produced during slow roll inflation with $r \sim 0.1$ \cite{Aasi:2014zwg},  the orange line depicts GWs from models such as axion inflation \cite{Cook:2011hg} that lead to enhanced GW production at frequencies $\sim 1$ Hz while remaining consistent with CMB bounds on $r$ (measured at $10^{-18}$ Hz), the blue line shows the expected GW strains produced by a network of cosmic strings with tension $G\mu \sim 10^{-16}$ \cite{DePies:2007bm} and the green curve (RS1) is an example of the spectrum that might be produced by a phase transition in the early universe \cite{RS1phaseTransition}.
		}
		\label{fig:stochastic sensitivity}
	\end{center}
\end{figure}

There are multiple possible uses for the resonant mode.  An important example is detection of a stochastic background of gravitational waves such as arises from inflation or other cosmological sources.  The stochastic strain sensitivity of the resonant detector is shown in Fig.~\ref{fig:stochastic sensitivity}.  The $1~\text{Hz}$ band may be a particularly promising band for detecting cosmological sources because it may have less ``noise" from astrophysical binaries \cite{Farmer:2003pa,Cutler2006}.  A satellite based detector may for example naturally allow a total interrogation time around 100 s, thus giving a broadband sensitivity down to $\sim 10^{-2}$ Hz.  It would be useful in this case to use resonance and give up sensitivity at those low frequencies where it is not useful, to gain sensitivity around 1 Hz.  A resonant enhancement of $Q \sim 100$ would move the peak sensitivity to 1 Hz and significantly boost the sensitivity to a cosmological signal. The resonant frequency can be changed by order one, thus allowing measurement of the shape of the spectrum for a stochastic signal, and for example possibly allowing measurement of the inflationary gravitational wave spectrum.  Improvements beyond the example sensitivity curve shown in Fig.~\ref{fig:stochastic sensitivity} could even possibly allow the detector to reach the level predicted by simple high-scale inflation models (with $r \sim 0.1$) \cite{Aasi:2014zwg}.

The detectability of these stochastic cosmological sources is subject to the GW background produced by unresolved black-hole mergers \cite{TheLIGOScientific:2016wyq}. A tunable, resonant detector with enhanced sensitivity might help resolve this background and deserves further study.

Another important use of resonance is for astrophysical signals such as inspiraling binaries.  At the lower frequencies around 1 Hz and below these binaries last a very long time and are thus rather monochromatic.  If a binary is detected, for example in broadband mode, resonant mode can be used to focus on it, greatly increasing the SNR in a much shorter time.  This would allow significantly more information to be learned about the binary including parameters such as masses, distance, and in particular the direction to the binary.  For both a satellite based detector and a terrestrial detector, the baseline will rotate (either with the earth or the satellite's motion), periodically changing the projection of the signal onto the detector baseline.  The direction of the source can be inferred by observing the resulting variation in signal amplitude, ideally at many points in time throughout the period of rotation of the detector.  The high SNR in resonant mode means that the source can be resolved with less averaging time, allowing a much more accurate measurement of the direction of the source than would be possible with a broadband detector.

The resonant mode not only helps gain information after a source has been detected, it also improves the detectability of long-lived sources.  The resonant frequency can be scanned over a broad bandwidth.  For example, if one chooses to operate in this swept-resonance mode, the strain sensitivity to such an astrophysical source is improved roughly as $\sim \sqrt{Q}$.  This is because the strain sensitivity on resonance is increased by $Q$, while the integration time spent in each particular frequency band is decreased by a factor $\sim Q$ but this only costs as the square root of time.  Thus the swept-resonance mode may have significant use for detecting astrophysical sources as well as for gaining information about them once they are detected.

\subsection{Sensitivity curve constraints}\label{Sec:SensitivityConstraints}

There are several practical constraints that determine the sensitivity shown in Fig.~\ref{fig:strain sensitivity}.  First, as mentioned above, there is a trade-off between LMT atom optics and resonant enhancement that limits the useful range of the ideal scaling $\propto n Q$.  In particular, due to imperfect transfer efficiency and other losses associated with the atom optics pulses, there is in practice a maximum number of pulses $n_\text{max}$ that can be applied during a single interferometer.  Here we assume $n_\text{max} = 10^3$, which is an order of magnitude above the current state-of-the-art for a discrete pulse LMT sequence\cite{54cm}.  For the resonantly enhanced LMT interferometer, the total number of pulses is $n_\text{p}=2Q(2n-1)+1$, so keeping $n_\text{p}\le n_\text{max}$ puts a constraint on the product of $n$ and $Q$ that we respect in Fig.~\ref{fig:strain sensitivity}.

Another practical constraint is that the total interferometer duration must be less than the lifetime of the atomic ensemble.  The maximum useful lifetime is limited by losses due to collisions with background gas atoms, scattering of stray light, the natural lifetime of the excited state, and expansion due to finite temperature.  For a resonant interferometer with $Q$ oscillations, the total time is constrained by $2T Q \le T_\text{max}$.  We assume a maximum time of $T_\text{max} = 300~\text{s}$.  This can be reached with modest vacuum requirements ($\lesssim 10^{-10}~\text{Torr}$) and with atom ensemble temperatures of $\sim 10~\text{pK}$, similar to what has already been demonstrated using delta kick cooling \cite{picokelvin}.  This constraint limits the resonant enhancement $Q$ at low frequencies and is responsible for the stair-step shape of the sensitivity curve in Fig.~\ref{fig:strain sensitivity}.

The $n_\text{max}$ and $T_\text{max}$ combined constraints explain the shape of the black sensitivity envelope in Fig.~\ref{fig:strain sensitivity}.  Respecting $T_\text{max}$, higher frequencies allow for larger $Q$, but for a fixed $n_\text{max}$ the maximum allowed $Q$ is smaller for higher values of $n$.  As a result, for each value of $n$ (red, green, blue, and orange shaded regions) the sensitivity improves at first going to higher frequencies, but at a certain point $n_\text{max}$ is reached and $Q$ can no longer increase.  This corresponds to the best sensitivity for that value of $n$, and at frequencies above this the sensitivity begins to degrade in accordance with Eq.~\ref{Eq:PeakPhaseShift}.  As a result, the peak sensitivity occurs at higher frequencies for smaller $n$.  While increasing $n$ improves the sensitivity scale factor, it reduces the maximum $Q$, so the net result is an increase in sensitivity at low frequency but a decrease at high frequency.  Increasing $n$ also limits the maximum allowed frequency, since the total duration of each LMT atom optic must be less than the interferometer time: $2(n-1)L/c < T$.  The upshot of these considerations is that different values of $n$ are preferable in different frequency ranges.  We emphasize that the $n$ and $Q$ parameters used in Fig.~\ref{fig:strain sensitivity} are only optimized for one particular physical design and would be different for other baseline lengths and telescope diameters.

\subsection{Noise constraints} \label{NoiseReqs}

The sensitivity curve in Fig.~\ref{fig:strain sensitivity} is only an atom shot noise limited curve.  It does not take into account all the other possible noise sources, since  the level of those noise sources depends on other parameters of the detector.  The purpose of this paper is to point out a new tool for gravitational wave detection.  It is beyond the scope of the paper to create a full detector design.  Instead, in this Section we will discuss several of the most relevant noise sources and simply give the requirements on the experimental design that would allow those noise sources to be below the atom shot noise.  We will not discuss how to implement such constraints in a realistic design.

An important consideration for long baselines is the photon shot noise of the laser link \cite{Hogan:2015xla}.  Shot noise leads to uncertainty in the phase comparison between the incoming reference laser pulses and the local oscillator laser, and these phase errors accumulate for each pulse in the atom interferometer sequence.  In Fig.~\ref{fig:strain sensitivity} we assume a $d=50~\text{cm}$ telescope diameter and ensure that the accumulated photon shot noise is less than the atom shot noise $\delta\phi_a = 10^{-5}~\text{rad}/\sqrt{\text{Hz}}$ at all frequencies shown.  This constraint puts limits on $Q$ that become relevant at high frequencies when the fact that $T \ll T_\text{max}$ would otherwise allow larger $Q$.

Using the analysis described in \cite{Hogan:2015xla}, the minimum required telescope diameter is
\begin{equation}
d=50~\text{cm}\,\Big(\!\tfrac{L}{4.4\cdot 10^7~\!\text{m}}\!\Big)^{\!\frac{2}{5}}  \Big(\!\tfrac{1~\!\text{W}}{P_t}\!\Big)^{\!\frac{1}{10}}  \Big(\!\tfrac{2~\!\text{Hz}/200}{f_R/n_\text{p}}\!\Big)^{\!\frac{1}{5}}  \Big(\!\tfrac{10^{-5}/\sqrt{\text{Hz}}}{\overline{\delta\phi}_a}\!\Big)^{\!\frac{2}{5}}
\end{equation}
where $P_t$ is the power transmitted from one end of the laser link and $f_R$ is the repetition rate of the detector.  As discussed in \cite{Hogan:2015xla}, multiple concurrent interferometers are needed to achieve the desired repetition rates.  To eliminate dead time \cite{PhysRevLett.116.183003}, atom cloud preparation and interferometer operation must be interleaved.

To alleviate photon shot noise constraints further at very long baselines, it can also be advantageous to trade-off the baseline length for LMT enhancement.  To achieve a fixed target strain sensitivity at shorter baseline $L$, the effective light path length $L_\text{eff} = n L$ of the LMT pulse sequence can be held constant by increasing $n$.  In this way, the LMT sequences have the effect of folding a long baseline into a smaller physical separation, much like how the Fabry-Perot resonators in the arms of LIGO give a longer effective baseline.  By reducing the physical separation, the light collection demands on the telescopes are reduced since the beams diverge less.  At a fixed level of photon shot noise, the minimum required telescope diameter scales as $d_\text{min}\sim L^{2/5} n^{1/5}\sim L_\text{eff}^{2/5} n^{-1/5}$ \cite{Hogan:2015xla}, which is reduced as $n$ in increased.  This scaling accounts for the fact that increasing $n$ has the unwanted side effect of increasing the total number of pulses, and thereby contributing additional photon shot noise to the detector.

The sensitivity curves in Fig.~\ref{fig:strain sensitivity} require tight constraints on the laser wavefront.  Laser wavefront aberrations $\delta\lambda/\lambda$ couple to satellite transverse position noise $\overline{\delta x}$ as discussed in \cite{Hogan:2015xla}.  Since the phase errors from aberrations add coherently with each pulse, sequences with $n_\text{p}$ near $n_\text{max}$ are the most challenging.  The wavefront requirement is
\begin{equation}
\delta\lambda=\tfrac{\lambda}{300} \, \Big(\!\tfrac{200}{n Q}\!\Big)\!\Big(\!\tfrac{\Lambda}{1~\text{cm}}\!\Big)\!\left(\!\tfrac{\overline{\delta\phi}_a}{10^{-5}~\text{rad}/\sqrt{\text{Hz}}}\!\right)\left(\!\tfrac{1~\text{nm}/\sqrt{\text{Hz}}}{\overline{\delta x}}\!\right)
\end{equation}
where $\Lambda$ is the aberration wavelength.  Wavefront requirements may be traded against constraints on the satellite position noise $\overline{\delta x}$.  It may be possible to relax these requirements to some degree by imaging the atom ensemble and resolving the phase imprint of the aberrations.

Timing jitter in the heterodyne laser link can result in residual sensitivity to laser frequency noise \cite{Hogan:2015xla}.  Assuming the pulses can be synchronized to ensure time delays less than $t_d \sim 1~\text{ns}$, the frequency noise amplitude spectral density of the interferometer laser must be \cite{Hogan:2015xla}
\begin{equation}
\overline{\delta\omega} =\, 2\pi\!\times\! 10~\!\tfrac{\text{Hz}}{\sqrt{\text{Hz}}}\,\, \Big(\!\tfrac{200}{n Q}\!\Big)\!\left(\!\tfrac{1~\text{ns}}{t_d}\!\right)\!\left(\!\tfrac{\overline{\delta\phi}_a}{10^{-5}~\text{rad}/\sqrt{\text{Hz}}}\!\right).
\end{equation}

Kinematic noise such as from acceleration noise of the laser platform can couple to the detector if there is a non-zero relative velocity $\Delta v$ between the two interferometers \cite{Graham:2012sy}.  For a satellite-based detector, relative velocity between the interferometers may arise because of perturbations to the orbits.  The platform acceleration noise limit for the sensitivity shown in Fig.~\ref{fig:strain sensitivity} is
\begin{equation}
\delta a =\, \!\tfrac{10^{-7} g}{\sqrt{\text{Hz}}}\,\, \Big(\!\tfrac{200}{n Q}\!\Big)\!\left(\!\tfrac{1~\text{cm/s}}{\Delta v}\!\right)\!\left(\!\tfrac{\overline{\delta\phi}_a}{10^{-5}~\text{rad}/\sqrt{\text{Hz}}}\!\right)\!\left(\!\tfrac{\omega_r}{2\pi\times 40~\text{mHz}}\!\right)^2
\end{equation}
where $g=9.8~\text{m/$\text{s}^2$}$ is Earth's gravity and $\omega_r$ is the target frequency of the detector.  Here we consider a relative velocity of $\Delta v\sim 1~\text{cm/s}$, but the actual velocity perturbations depend strongly on the details of the orbit.  Although platform acceleration noise is magnified for sequences with many laser pulses, this is somewhat compensated by the fact that the acceleration requirement scales favorably with increasing frequency.

\section{Conclusions}

The resonant mode enhances the capabilities of previously proposed gravitational wave detectors based on atomic sensors.  The same instrument could be run either in broadband or resonant mode, and the switch can be made in real time since it only requires changing the laser pulse sequence.  This mode is a new tool for these detectors that can provide design flexibility.  For example, it can be used to accommodate constraints on the maximum size of the interferometer region, allowing a significantly more sensitive detector to fit in a confined region, such as inside a satellite.

The sensitivity to astrophysical sources and the information gained about them can be enhanced with the resonant mode of operation.  For instance, it may significantly boost the precision of the direction measurement.  Additionally, a specific binary may be followed in frequency until it reaches the LIGO band above 10~Hz, improving the confidence of the detection and the measurement of the binary's parameters by observing it in multiple frequency bands.

The resonant mode may significantly enhance the use of these gravitational wave detectors for cosmology.  A satellite-based detector operating in the resonant mode could have significantly enhanced sensitivity around 1 Hz, which may be the optimal band to search for cosmological sources such as inflation.  The only known ways to directly observe such signals from inflation are the CMB \cite{Kamionkowski1999} and direct gravitational wave detection in this frequency band.  The 1 Hz band probes a very different part of the inflationary epoch than the CMB, and a different part of the inflaton potential.  In the ideal scenario, a detection in both the CMB and a direct GW detector would provide a powerful test of the inflationary spectrum over roughly 18 orders of magnitude in frequency, probing details of the mechanism behind inflation.  Such a long lever-arm would allow precise measurement of inflationary parameters that may not otherwise be observable.  Gravitational waves offer possibly the only way to directly observe the universe before last scattering, potentially probing energy scales far above laboratory experiments.

\section*{Acknowledgements}
We would like to thank S.~Dimopoulos, R.~Flauger,  S.~Kolkowitz, M.~Lukin, J.~Mardon, P.~Michelson, B.~Saif, L.~Senatore, R.~Walsworth and J.~Ye.
PWG acknowledges the support of NSF grant PHY-1316706, DOE Early Career Award DE-SC0012012, and the W.M.~Keck Foundation.  
SR was supported in part by the NSF under grants PHY-1417295 and PHY-1507160, the Simons Foundation Award 378243. 
This work was supported in part by the Heising-Simons Foundation grants 2015-037 and 2015-038.

\end{document}